\documentclass[11pt,onecolumn,amssymb,nofootinbib]{revtex4}
\usepackage{amsmath, amsthm, amscd, amssymb}
\usepackage{bm}
\usepackage{bbm}

\begin{document}

\title{\bf Quantized Gauged Massless Rarita-Schwinger Fields}

\author{Stephen L. Adler}
\email{adler@ias.edu} \affiliation{Institute for Advanced Study,
Einstein Drive, Princeton, NJ 08540, USA.}

\begin{abstract}
We study quantization of a minimally gauged massless Rarita-Schwinger field, by both Dirac bracket and
functional integral methods.  The Dirac bracket approach in covariant radiation gauge leads to an
anticommutator that has a non-singular  limit as gauge fields approach zero,  is manifestly positive semidefinite, and is Lorentz
invariant.   The constraints also have the form needed to apply the Faddeev-Popov method for deriving a
functional integral, using  the same constrained Hamiltonian and inverse constraint matrix that appear in
the Dirac bracket approach.

\end{abstract}

\maketitle

\section{Introduction}

In this paper we continue the study of gauging a massless Rarita-Schwinger field begun in the preceding paper \cite{adlerclassical},
referred to henceforth as (I) ,  which dealt principally  with the classical case  (with a small excursion into first quantization).  Here we turn
to a detailed examination of quantization of a gauged massless Rarita-Schwinger field.  Our main aim is to show that a consistent quantization
is possible in gauge covariant radiation gauge, avoiding the problem of non-positivity of the canonical anticommutator first noted by Johnson
and Sudarshan \cite{johnson} and later rederived by Velo and Zwanziger \cite{velo}. Other objections to gauging a massless Rarita-Schwinger field -- the issue  of superluminal signaling, and various ``on-shell no-go''theorems -- have already been taken up in (I).  In referring to a formula numbered ``Eq. (\#)'' in the preceding paper we shall use the notation
``Eq. (I-\#)'' ,  while non-hyphenated equation numbers refer to equations from this paper.

In Sec. 2  we give the Hamiltonian form of the
equations of motion and constraints, and introduce the Dirac bracket. This can be done without imposing a gauge fixing condition; in particular, we do not use the condition $\Psi_0=0$ that was imposed in an initial arXiv posting \cite{adlerarxiv} of this paper.
When a gauge fixing condition is omitted, the equation of motion for $\vec \Psi$ computed from the Dirac bracket agrees with the equation of motion of Eq. (I-29), in the form obtained when $\Psi_0$ is eliminated by using the secondary constraint
$\omega=0$; this demonstrates that the Dirac formalism is working correctly in the Rarita-Schwinger equation context.
However, in the absence of a gauge fixing constraint, the Dirac bracket anticommutator of $\vec \Psi$ with ${\vec \Psi}^{\dagger}$  agrees with the anticommutator calculated in \cite{johnson} and \cite{velo}, which is singular in the limit of vanishing gauge fields and
is not positive semidefinite.

In Sec. 3  we study the Dirac bracket in its classical and quantum forms with imposition of a covariant radiation gauge constraint. We show that now  the quantum Dirac bracket has the requisite positivity
properties to be an anticommutator; related details are given in Appendix A. In Sec. 4 we give an alternative approach to proving positivity
of the anticommutator in covariant radiation gauge, based on writing a Lagrangian for the equation of motion for $\vec \Psi$ in which $\Psi_0$ has
already been eliminated by use of  the
secondary constraint.  In Sec. 5 we discuss Lorentz covariance of covariant radiation gauge and show Lorentz invariance of the
Dirac bracket. In Sec. 6 we turn to path integral quantization in covariant radiation gauge, leading to a formalism closely resembling the Dirac bracket approach. A brief concluding discussion is given in Sec. 7.

Our conclusion from this paper and the preceding one  is that one can consistently gauge a massless Rarita-Schwinger field, at both the classical
and quantum levels.   This opens the possibility of using gauged Rarita-Schwinger fields as part of the anomaly cancelation mechanism in grand unified models, with anomalies of the spin $\frac{1}{2}$ fields canceling against the
spin $\frac{3}{2}$ anomaly.

\section{Hamiltonian form of the equations and the Dirac bracket}
The standard route to canonical quantization is to  transform the Lagrangian equations to Hamiltonian form, and to take the constraints into
account by replacing the classical brackets by Dirac brackets.  In carrying this out, we will simplify the formulas by
making the gauge choice $A_0=0$ for the non-Abelian gauge fields.  This gauge choice is always attainable, and leaves a residual non-Abelian
gauge invariance with time-independent gauge parameter.   The Hamiltonian will then be covariant with respect to this restricted gauge transformation.  For the moment, in discussing the canonical Hamiltonian and bracket formalism,  we will allow $\vec A$ to be time dependent, so that $\vec E \ne 0$.  But when we turn to the Dirac bracket construction corresponding to a constrained Hamiltonian, which is simplest in the case of time-independent constraints,  we will assume a time-independent $\vec A$, corresponding in $A_0=0$ gauge to $\vec E=0$.   (If we carry along the $A_0$  term in the formulas then time-independent fields would not require $\vec E=0$.  So this specialization can be avoided at the price of somewhat lengthier equations.)

From the action $S(\Psi_{\mu})=\int dt L(\Psi_{\mu})$  of Eq. (I-23) and the canonical momentum $\vec P = \frac{1}{2} \vec{\Psi}^{\dagger} \times \vec \sigma$, we find the canonical Hamiltonian to be
\begin{align}\label{eq:ham}
H=&\int d^3x \partial_0 \vec \Psi \cdot \vec P    - L~~~\cr
=-&\frac{1}{2}\int d^3x  [-\Psi_{0}^{\dagger} \vec \sigma \cdot \vec D \times \vec {\Psi}
+\vec {\Psi}^{\dagger} \cdot \vec \sigma \times \vec D \Psi_{0}
+\vec{\Psi} ^{\dagger} \cdot \vec D \times \vec \Psi ]~~~\cr
=-&\frac{1}{2}\int d^3x  [-\Psi_{0}^{\dagger} \vec \sigma \cdot \vec D \times \vec {\Psi}
+(i\vec P-\vec P \times \vec \sigma) \cdot (\vec \sigma \times \vec D \Psi_{0}
+\vec D \times \vec \Psi) ]~~~,\cr
\end{align}
where in the final line we have used the inversion formula $\vec {\Psi}^{\dagger}=i\vec P -\vec P \times \vec \sigma$.

We can now compute the classical brackets of various quantities with $H$.  From
\begin{align}\label{eq:psieqmo}
\frac{d\vec \Psi}{dt}=&[\vec \Psi, H]_C=\frac{1}{2}[i(\vec \sigma \times \vec D \Psi_0+\vec D\times \vec \Psi)-\vec \sigma \times (\vec \sigma \times \vec D \Psi_0+\vec D\times \vec \Psi)]\cr
=&\vec D \Psi_0+ \frac{1}{2} [-\vec \sigma \times (\vec D \times \vec{\Psi}) + i \vec D \times \vec \Psi]~~~,\cr
\end{align}
we obtain the $\vec \Psi$ equation of motion in the form given in Eq. (I-29).  Similarly,
from the bracket of $\vec P$ with $H$ we find the equation of motion for $\vec{\Psi}^{\dagger}$.
Turning to brackets of the constraints with $H$, starting with $P_{\Psi_0^{\dagger}}$,   we find
\begin{equation}\label{eq:P0dagbrac}
\frac{dP_{\Psi_0^{\dagger}}}{dt}=[P_{\Psi_0^{\dagger}},H]_C=-\frac{1}{2}\chi~~~,
\end{equation}
and so $P_{\Psi_0^{\dagger}}=0$ for all times implies that $\chi=0$.  For the total time derivative
of $\chi$, we have
\begin{equation}\label{eq:chibrac}
\frac{d\chi}{dt}=\frac{\partial \chi}{\partial t}+[\chi,H]_C=\vec \sigma \times g\frac{\partial  \vec A}{\partial t}\cdot \vec{\Psi}+ [\chi,H]_C=-ig \omega~~~,
\end{equation}
and so $\chi=0$ for all times implies that $\omega$ defined in Eq. (I-28) vanishes.  Since $\omega$ contains
a term proportional to $\Psi_0$, to continue this process by calculating the time derivative of $\omega$, we must obtain
$d\Psi_0/dt$ from a bracket of $\Psi_0$ with $H$ (and similarly for $d\Psi_0^{\dagger}/dt)$.  This requires adding to $H$ a term
\begin{equation}\label{eq:deltah}
\Delta H=-\int d^3x \left[P_{\Psi_0} \frac {d\Psi_0}{dt} + P_{\Psi_0^{\dagger}}\frac{d\Psi_0^{\dagger}}{dt}\right]~~~.
\end{equation}
Requiring $\Delta H$ to be self-adjoint then imposes the requirement
\begin{equation}\label{eq:Padjoint}
P_{\Psi_0}^{\dagger}=- P_{\Psi_0^{\dagger}}~~~,
\end{equation}
which was noted following Eq. (I-56).  As noted in (I), the chain of successive brackets with $H$ starting from
$P_{\Psi_0^{\dagger}}$ and continuing to $\chi,\omega,...$ leads only to constraints involving $\vec \Psi$ and $\Psi_0$ but never
their adjoints.  The doubling of the set of constraints, which turns the first class constraints into second class ones, comes
from requiring that the adjoint of each fermionic constraint also be a constraint, not from taking successive brackets with $H$.

We are now ready to implement the Dirac bracket procedure.  The basic idea is to change the canonical bracket $[F,G]_C$ to a modified
bracket $[F,G]_D$, which projects $F$ and $G$ onto the subspace obeying the constraints, so that the constraints are built into
the brackets, or after quantization, into the canonical anticommutators.  The constraints  can then be``strongly'' implemented
in the Hamiltonian by setting terms proportional to the constraints to zero.  After integration by parts the second line of Eq. \eqref{eq:ham} takes the form
\begin{equation}\label{eq:ham1}
H=-\frac{1}{2}\int d^3x  [-\Psi_{0}^{\dagger} \chi - \chi^{\dagger}  \Psi_{0}
+\vec{\Psi} ^{\dagger} \cdot \vec D \times \vec \Psi ]~~~,
\end{equation}
so setting the  constraints $\chi^{\dagger},\, \chi$ respectively  to zero in Eq. \eqref{eq:ham1},
we see that the constrained Hamiltonian is just
\begin{align}\label{eq:ham2}
H=&-\frac{1}{2}\int d^3x
\vec{\Psi} ^{\dagger} \cdot \vec D \times \vec \Psi \cr
=&-\frac{1}{2}\int d^3x
(i \vec P - \vec P \times \vec \sigma) \cdot \vec D \times \vec \Psi \cr
\end{align}
which coincides with the energy integral computed in Eq. (I-35) from the stress-energy tensor.

We proceed now to calculate the Dirac bracket for the case when $F=F(\vec \Psi)$ and $G=G(\vec \Psi, \vec \Psi^{\dagger}
)$; the case
when $F=F(\vec \Psi^{\dagger})$ can then be obtained by taking the adjoint, and the case when  $F=F(\vec \Psi, \vec \Psi^{\dagger})$
can be obtained by combining the extra bracket terms from both calculations. When $F$  has no dependence on $\vec \Psi^{\dagger}$,
it has vanishing brackets with the constraints $\phi_a$ of Eq. (I-55) and nonvanishing brackets with the constraints
$\chi_a$ of Eq. (I-56).  The Dirac bracket then has the form \big(see Eqs. (I-A20) and (I-A21) for why $M^{-1}$ appears\big)
\begin{equation}\label{eq:dirac1}
[F,G]_D=[F,G]_C-\sum_a\sum_b [F,\chi_a]_C M^{-1}_{ab} [\phi_b,G]~~~,
\end{equation}
where $M_{ab}(\vec x, \vec y) =  [\phi_a(\vec x),\chi_b(\vec y)]_C$ is the matrix defined in
Eqs. (I-58) and (I-59). We recall that this matrix has the form
\begin{equation}\label{eq:pstructure1}
M=\left( \begin{array} {cccc}
 0&-1&0&0 \\
 1&{\cal U}&{\cal S}&{\cal T} \\
 0&{\cal V}&{\cal A}&{\cal B} \\
 0&{\cal W}&{\cal C}&{\cal D} \\
 \end{array} \right)~~~,
\end{equation}
where in the $SU(n)$ gauge field case, each entry in $M$ is a $2n\times 2n $ matrix.  Using the block inversion
method given in Eqs. (I-A18) and (I-A19), we find that $M^{-1}$ is given by
\begin{equation}\label{eq:minverse}
M^{-1}=\left( \begin{array} {cccc}
 \Sigma&1&-({\cal S}\cal{F}+{\cal T}{\cal H})&-({\cal S}{\cal G}+{\cal T}\cal{I}) \\
 ~-1~~~&~~~0~~~&~~~0~~~&~~~0~~~ \\
 {\cal F}{\cal V}+{\cal G}{\cal W}&0&{\cal F}&{\cal G} \\
 {\cal H}{\cal V}+{\cal I}{\cal W}&0&{\cal H}&{\cal I} \\
 \end{array} \right)~~~,
\end{equation}
where
\begin{equation}\label{eq:sigmadef}
\Sigma={\cal U}-{\cal S}( {\cal F}{\cal V}+{\cal G}{\cal W})-{\cal T}( {\cal H}{\cal V}+{\cal I}{\cal W} )~~~,
\end{equation}
and where ${\cal F}$, ${\cal G}$, ${\cal H}$, ${\cal I}$ are the elements of the block inversion of the matrix
$N$ of Eq. (I-60),
\begin{equation}
\left(\begin{array} {cc}
  {\cal F}&{\cal G} \\
 {\cal H}&{\cal I} \\
 \end{array} \right)  \left(\begin{array} {cc}
  {\cal A}&{\cal B} \\
 {\cal C}&{\cal D} \\
 \end{array} \right)
=\left(\begin{array} {cc}
  1&0 \\
 0&1 \\
 \end{array} \right)  ~~~.
\end{equation}
Substituting these into Eq. \eqref{eq:dirac1} we find for the Dirac bracket a lengthy expression, which
simplifies considerably after noting that $[F(\vec \Psi),\chi_1]_C=[F(\vec \Psi), -P_{\Psi_0}]_C=0$ and
$[\phi_1,G(\vec \Psi, \vec{\Psi}^{\dagger})]_C=[P_{\Psi_0^{\dagger}},G(\vec \Psi, \vec{\Psi}^{\dagger})]_C=0$,
leaving the relatively simple formula
\begin{align}\label{eq:simplebrac}
[F(\vec \Psi),G(\vec \Psi, \vec{\Psi}^{\dagger})]_D=& [F(\vec \Psi),G(\vec \Psi, \vec{\Psi}^{\dagger})]_C   \cr
-& [F(\vec \Psi),\chi_3]_C \Big({\cal F}\, [\phi_3,G(\vec \Psi, \vec{\Psi}^{\dagger})]_C + {\cal G} \,[\phi_4,G(\vec \Psi, \vec{\Psi}^{\dagger})]_C\Big)   \cr
-& [F(\vec \Psi),\chi_4]_C \Big({\cal H}\, [\phi_3,G(\vec \Psi, \vec{\Psi}^{\dagger})]_C + {\cal I}\, [\phi_4,G(\vec \Psi, \vec{\Psi}^{\dagger})]_C\Big) ~~~.  \cr
\end{align}
We note that only the matrix $N$ enters, in this case through its inverse, rather than the full
matrix of constraint brackets $M$.  The final step is to evaluate the inverse block matrix elements ${\cal F},\,{\cal G},\,{\cal H},\,{\cal I}$ from
the expressions for ${\cal A},\,{\cal B},\,{\cal C},\,{\cal D}$, again by using the block inversion formulas of Eqs. (I-A18) and (I-A19).
Let us define the Green's function ${\cal D}^{-1}(\vec x-\vec y)$
by
\begin{equation}\label{eq:green}
\big( i(\vec L_{\vec x})^2 + \vec \sigma \cdot \vec L_{\vec x} \times \vec L_{\vec x} \big){\cal D}^{-1}(\vec x-\vec y)=\delta^3(\vec x-\vec y)~~~,
\end{equation}
and a second Green's function ${\cal Z}(\vec x - \vec y)$ by
\begin{align}\label{eq:greenz}
{\cal Z}(\vec x - \vec y)=&{\cal A} - {\cal B} {\cal D}^{-1} {\cal C}\cr
=&-2ig \vec  \sigma \cdot \vec B \delta^3(\vec x-\vec y) - 4 \vec D_{\vec x} \cdot \vec L_{\vec x} {\cal D}^{-1}(\vec x-\vec y) \vec{L}_{\vec y} \cdot \overleftarrow{D}_{\vec y}~~~.\cr
\end{align}
where in covariant radiation gauge $\vec L=\vec D$.
Then the needed inverse block matrices are
\begin{align}\label{eq:inverseblock}
{\cal F}=&{\cal Z}^{-1}~~~,\cr
{\cal G}=&-{\cal Z}^{-1}{\cal B}{\cal D}^{-1}~~~,\cr
{\cal H}=&-{\cal D}^{-1}{\cal C}{\cal Z}^{-1}~~~,\cr
{\cal I}=&{\cal D}^{-1} + {\cal D}^{-1} {\cal C} {\cal Z}
^{-1} {\cal B} {\cal D}^{-1}~~~.\cr
\end{align}

We wish now to apply the Dirac bracket formula to the cases (i)  $F(\vec \Psi)=\vec \Psi$ and $G(\vec \Psi, \vec{\Psi}^{\dagger})=\vec{\Psi}^{\dagger}$, and
(ii) $F(\vec \Psi)=\vec \Psi$ and $G(\vec \Psi, \vec{\Psi}^{\dagger})=H$, with $H$ the constrained Hamiltonian of Eq. \eqref{eq:ham2}.
The following canonical brackets are needed for this:
\begin{align}\label{eq:neededbracs}
[\vec \Psi(\vec x),\chi_3(\vec y)]_C=&2 \vec D_{\vec x} \delta^3(\vec x-\vec y)~~~,\cr
[\vec \Psi(\vec x),\chi_4(\vec y)]_C=&(i \vec L_{\vec x}- \vec \sigma \times \vec L_{\vec x})\delta^3(\vec x-\vec y)~~~,\cr
[\phi_3(\vec x),\vec{\Psi}^{\dagger}(\vec y)]_C=&2 \vec D_{\vec x} \delta^3(\vec x-\vec y)= -2 \delta^3(\vec x-\vec y) \overleftarrow{D}_{\vec y}~~~,\cr
[\phi_4(\vec x),\vec{\Psi}^{\dagger}(\vec y)]_C=&-(i \vec L_{\vec x}-  \vec L_{\vec x} \times  \vec \sigma) \delta^3(\vec x-\vec y)
=\delta^3(\vec x-\vec y)(i \overleftarrow{L}_{\vec y}- \overleftarrow{L}_{\vec y} \times \vec \sigma )~~~,\cr
[\phi_3(\vec x),H]_C=& ig\vec B(\vec x) \cdot \vec \Psi(\vec x)~~~,\cr
[\phi_4(\vec x),H]_C=&\frac{1}{2} (i\vec L_{\vec x}- \vec L_{\vec x} \times \vec \sigma) \times \vec D_{\vec x} \cdot \vec \Psi(\vec x)~~~. \cr
\end{align}
Additionally, for case (i) we need the canonical bracket
\begin{align}\label{eq:psipsibarbrac}
[\Psi_i(\vec x),\Psi^{\dagger}_j(\vec y)]_C= &[\Psi_i(\vec x),iP_j(\vec y)-\epsilon_{jkl}P_k(\vec y)\sigma_l]_C\cr
= &-i (\delta_{ij}+i\epsilon_{jil}\sigma_l)\delta^3(\vec x-\vec y)=-i \sigma_j\sigma_i \delta^3(\vec x-\vec y)=-2i \Big(\delta_{ij}-\frac{1}{2}\sigma_i\sigma_j\Big)\delta^3(\vec x-\vec y)  ~~~,\cr
\end{align}
and for case (ii) we need the canonical bracket
\begin{equation}\label{eq:psiHbrac}
[\Psi_i(\vec x),H]_C=\frac{1}{2}\Big(i \vec D_{\vec x} \times \vec \Psi(\vec x) - \vec \sigma \times \big(\vec D_{\vec x} \times \vec \Psi(\vec x)\big)\Big)_i~~~.
\end{equation}

Up to this point, we have not specialized  $\vec L$  so as to make it easy to ascertain what the formulas become when gauge fixing is omitted (as in \cite{johnson} and \cite{velo}). When $\vec L=0$, the matrix $N$ degenerates to its upper left element ${\cal A}$.  This is reflected in the fact that  ${\cal Z}$ of Eq. \eqref{eq:greenz} simplifies to
\begin{equation}\label{eq:zsimp}
{\cal Z}(\vec x - \vec y)={\cal A} =-2ig \vec  \sigma \cdot \vec B \delta^3(\vec x-\vec y)~~~,
\end{equation}
which is a local function of $\vec x$ and so is algebraically invertible.
The Dirac bracket of $\vec \Psi(\vec x)$ with the constrained Hamiltonian now simplifies to
\begin{align}\label{eq:psihambrac0}
\frac{d\vec \Psi(\vec x)}{dt}=& [\vec \Psi(\vec x),H]_D=\frac{1}{2}[i\vec D_{\vec x} \times \vec \Psi(\vec x)-\vec \sigma \times \big(\vec D_{\vec x} \times \vec \Psi(\vec x) \big)]
-\int d^3y\Big\{ 2 \vec D_{\vec x}\Big[{\cal Z}^{-1}(\vec x-\vec y) i g \vec B(\vec y) \cdot \vec \Psi(\vec y)\Big]\Big\}\cr
=& \frac{1}{2}[i\vec D_{\vec x} \times \vec \Psi(\vec x)-\vec \sigma \times \big(\vec D_{\vec x} \times \vec \Psi(\vec x) \big)]
+\vec D_{\vec x} \frac{1}{\vec \sigma \cdot \vec B(\vec x)} \vec B(\vec x) \cdot \vec \Psi(\vec x)
~~~.\cr
\end{align}
The second line of this equation is just the $\vec \Psi$ equation of motion in the form of Eq. (I-29)  (when $A_0=0$), with $\Psi_0$ eliminated
by using the secondary constraint, which when $\vec E=0$ reads $\vec \sigma \cdot \vec B \Psi_0= \vec B \cdot \vec \Psi$. This shows
that the Dirac bracket formalism correctly incorporates the $\Psi_0$ term of Eq. (I-29).  The reason a local result is obtained from this
calculation is that in the absence of gauge fixing, the Dirac bracket only projects into the subspace that preserves the
primary constraint $\chi=0$, and since the equation of motion of Eq. (I-29) preserves this constraint, it already resides in the
subspace projected into by the $\vec L=0$ Dirac bracket.

When $\vec L=0$, for the Dirac bracket of $\vec \Psi_i(\vec x)$ with $\vec \Psi^{\dagger}_j(\vec y)$ we find
\begin{align}\label{eq:velobrac}
[\Psi_i(\vec x),\Psi^{\dagger}_j(\vec y)]_D= &[\Psi_i(\vec x),\Psi^{\dagger}_j(\vec y)]_C-\int d^3w d^3z [\Psi_i(\vec x),\chi_3(\vec w)]_C {\cal Z}^{-1}(\vec w-\vec z) [\phi_3(\vec z),\Psi^{\dagger}_j(\vec y)]_C\cr
=&-2i\Big[ \Big(\delta_{ij}-\frac{1}{2}\sigma_i\sigma_j\Big)\delta^3(\vec x-\vec y)-D_{\vec x \,i}\frac{\delta^3(\vec x-\vec y)}{g \vec \sigma \cdot \vec B(\vec x)}\overleftarrow{D}_{\vec y\,j}\Big]\cr
=&-2i\langle \vec x|\Big[ \Big(\delta_{ij}-\frac{1}{2}\sigma_i\sigma_j\Big)1+\Pi_i\frac{1}{g \vec \sigma \cdot \vec B}\Pi_j\Big]|\vec y \rangle~~~,\cr
\end{align}
where in the final line we have written $iD_{\vec x\, i}=\Pi_i$ to relate to the abstract operator notation of Velo and Zwanziger \cite{velo}.  Multiplying the final line by $i$ to
convert the Dirac bracket to an anticommutator, and by a factor $1/2$ reflecting our different field normalization, Eq. \eqref{eq:velobrac} becomes  the expression for the
anticommutator given in the zero mass limit of Eq. (4.12) of \cite{velo}.  Using identities in Appendix A of (I), one can verify (as in Appendix C of \cite{velo}) that
\begin{equation}\label{eq:chiproj}
(\vec \sigma \times \vec D_{\vec x})_i \Big[ \Big(\delta_{ij}-\frac{1}{2}\sigma_i\sigma_j\Big)\delta^3(\vec x-\vec y)-D_{\vec x\,i}\frac{\delta^3(\vec x-\vec y)}{g \vec \sigma \cdot \vec B(\vec x)}\overleftarrow{D}_{\vec y\,j}\Big]=0~~~,
\end{equation}
that is, the constraint $\chi$ is explicitly projected to zero.  However, as noted in the Introduction to (I), the anticommutator of Eq.
\eqref{eq:velobrac} becomes singular as $\vec B \to 0$, rather than limiting to the free Rarita-Scwhinger anticommutator.  This problem
is a direct consequence of omitting a gauge-fixing constraint, by taking  $\vec L=0$ in calculating the matrix $N$.

Now setting $\vec L=\vec D$ for covariant radiation gauge, we find for the Dirac bracket of $\Psi_i(\vec x)$  with the constrained Hamiltonian,
\begin{align}\label{eq:psihambrac}
\frac{d\vec \Psi(\vec x)}{dt}=& [\vec \Psi(\vec x),H]
_D=\frac{1}{2}[i\vec D_{\vec x} \times \vec \Psi(\vec x)-\vec \sigma \times \big(\vec D_{\vec x} \times \vec \Psi(\vec x) \big)]\cr
-&\int d^3y\Big\{ 2 \vec D_{\vec x}\Big[{\cal F}(\vec x-\vec y) i g \vec B(\vec y) \cdot \vec \Psi(\vec y)\cr
+&{\cal G}(\vec x-\vec y)
 \frac{1}{2}\big(i \vec D_{\vec y}-\vec D_{\vec y} \times \vec \sigma\big) \times \vec D_{\vec y} \cdot \vec \Psi(\vec y)\Big]  \cr
+&(i \vec D_{\vec x} - \vec \sigma \times \vec L_{\vec x}) \Big[{\cal H}(\vec x-\vec y) i g \vec B(\vec y) \cdot \vec \Psi(\vec y)\cr
+&{\cal I}(\vec x-\vec y) \frac{1}{2}\big(i \vec D_{\vec y}-\vec D_{\vec y} \times \vec \sigma\big) \times \vec D_{\vec y} \cdot \vec \Psi(\vec y)\Big]\Big\}~~~.
\end{align}
The first line of this equation  gives the second term of the unconstrained equation of motion in the form of Eq. (I-29), while the remaining terms replace the first term of Eq. (I-29) to guarantee that
\begin{align}\label{eq:chivanish}
\frac{d\phi_3}{dt}=&\frac{d\chi}{dt}= \frac{d(\vec \sigma \times \vec D \cdot \vec \Psi)}{dt}=\sigma \times \vec D \cdot \frac{d\vec \Psi}{dt}=0~~~,\cr
\frac{d\phi_4}{dt}=&d\frac{\vec D \cdot \vec \Psi}{dt}=\vec D \cdot  \frac{d\vec \Psi}{dt}=0~~~,\cr
\end{align}
where we have used the fact that we are assuming that $\vec D$ is  time independent. That is, the Dirac bracket simultaneously projects the
equation of motion into the subspace where both $\chi=0$ and $\vec D \cdot \vec \Psi=0$.   The restriction to $\vec D$ time independent can be avoided by treating the gauge fields
as dynamical variables, taking into account their own constraint structure, and noting that the radiation gauge fixing constraint $\vec \nabla \cdot \vec P_{\vec A}=0$,
with $\vec P_{\vec A}$ the canonical momentum conjugate to $\vec A$, has nonvanishing fermionic  brackets with all Rarita-Schwinger constraints involving $\vec D=\vec \nabla + g \vec A$.  This requires an extension of the Dirac bracket construction to take the new, Grassmann-odd, brackets into account, and the extended Dirac bracket structure will then
obey Eq. \eqref{eq:chivanish} without requiring the assumption of a time independent $\vec A$ and $\vec D$.

With $\vec L=\vec D$, we find for the Dirac bracket of $\Psi_i(\vec x)$ with $\Psi_j^{\dagger}(\vec y)$,
\begin{align}\label{eq:psipsidagbrac}
[\Psi_i(\vec x),\Psi_j^{\dagger}(\vec y)]_D=&-2i \Big(\delta_{ij}-\frac{1}{2}\sigma_i\sigma_j\Big)\delta^3(\vec x-\vec y) \cr
+& 4\vec  D_{\vec x\, i}{\cal F}(\vec x-\vec y) \overleftarrow{D}_{\vec y\,j} -2  D_{\vec x\, i} {\cal G}(\vec x-\vec y)
(i\overleftarrow{D}_{\vec y}- \overleftarrow{D}_{\vec y} \times \vec \sigma)_j\cr
+&2(i\vec  D_{\vec x}- \vec \sigma \times \vec  D_{\vec x})_i {\cal H}(\vec x-\vec y) \overleftarrow{D}_{\vec y\, j}-(i\vec  D_{\vec x}
- \vec \sigma \times \vec  D_{\vec x})_i{\cal I}(\vec x -\vec y)  (i\overleftarrow{D}_{\vec y}- \overleftarrow{D}_{\vec y} \times \vec \sigma )_j~~~,\cr
\end{align}
which gives the generalization of Eq. \eqref{eq:velobrac} to the case when a covariant gauge fixing constraint is imposed.  This equation
will be further analyzed in the next section.

\section{Quantization of the anticommutator derived from the Dirac bracket and positivity  in covariant radiation gauge}

Given the Dirac bracket, the next step is to quantize, by multiplying all Dirac brackets by $i$ and then reinterpreting them as anticommutators or commutators of operators.  In the
case considered here, this can be done in a constructive way, as follows.  First let us replace the set of $2n$ component column vector constraints $\phi_a$ and $2n$ component row vector constraints $\chi_a$ by the set of $4n$ scalars given by their individual matrix elements.  Moreover, since the $\chi_a$ are the adjoints of the $\phi_a$, we can take linear combinations to make
all of these scalars self-adjoint.  Labeling the set of self-adjoint scalar constraints by $\Phi_a$, the Dirac bracket construction for the bracket of $F$ with $G$ reads
\begin{align}\label{eq:fullbrac}
[F,G]_D=&[F,G]_C-\sum_a\sum_b [F,\Phi_a]_C T^{-1}_{ab} [\Phi_b,G]_C~~~,\cr
T_{ab}=&[\Phi_a,\Phi_b]_C~~~,  \cr
\end{align}
with the matrix $T$ real.

We now observe that since the $\Phi_a$ are all {\it linear} in the scalar components of $\vec \Psi$ and  $ \vec \Psi^{\dagger}$, if we make the replacement $i[~,~]_C \to \{~,~\}_C$, with $\{~,~\}
$ the anticommutatior, and replace all Grassmann variables $\vec \Psi$ and $\vec \Psi^{\dagger}$ with operator variables having
the standard canonical anticommutators, then since there is no other operator structure the same real matrix $T_{ab}$ will be obtained.  Moreover,
if $F$ and $G$ are both linear in the scalar components of $\vec \Psi$ and  $\vec \Psi^{\dagger}$, the Grassmann bracket  $i[F,G]_C$ formed from scalar components of $F$ and $G$ will agree  with the canonical anticommutator  $i\{F,G\}_C$ formed from the corresponding operator scalar components, and will be a $c$-number.   Thus, for linear
$F$ and $G$ we can define a ``Dirac anticommutator" $\{F,G\}_D$  by
\begin{align}\label{eq:anticommbrac}
\{F,G\}_D=&\{F,G\}_C-\sum_a\sum_b \{F,\Phi_a\}_C T^{-1}_{ab} \{\Phi_b,G\}_C~~~,\cr
T_{ab}=&\{\Phi_a,\Phi_b\}_C~~~.\cr
\end{align}
When one or both of $F$ and $G$ is bilinear, the Grassmann bracket $i[F,G]_C$ formed from the scalar components of $F$ and $G$ will agree with the
canonical commutator formed from the corresponding operator scalar components, and we can  define a ``Dirac commutator'' by  a formula analogous to Eq. \eqref{eq:anticommbrac} in which each anticommutator with at least one bilinear argument is replaced by a commutator.  In this way  we get
a mapping of classical brackets into quantum anticommutators and commutators, that inherits  the algebraic properties of the Dirac bracket, including the
chain rule, with the Jacobi identities for odd and even Grassmann variables mapping to the corresponding   anticommutator and commutator Jacobi identities.

To complete this correspondence, we must show that the Dirac anticommutator of $\Psi_i^{\alpha\,u}$ and $\Psi_j^{\dagger \, \beta \,v}$ (with $\alpha =1,2,~\beta=1,2$ the spin indices, $u=1,...,n,~v=1,...,n$ the internal symmetry indices, and $i=1,2,3,~j=1,2,3$ the spatial vector
 indices) has the expected positivity properties of an
operator anticommutator, by showing that for an arbitrary set of complex functions $A_i^{\alpha\,u}(\vec x)$, we have
\begin{equation}\label{eq:pos1}
\int d^3x d^3y A_i^{\alpha\,u}(\vec x) A_j^{*\beta\,v}(\vec y) \{\Psi_i^{\alpha\,u}(\vec x),\Psi_j^{\dagger \, \beta\,v}(\vec y)\}_D \geq 0~~~.
\end{equation}
We demonstrate this in several steps, in covariant radiation gauge.  First we examine the conditions for positivity of the canonical anticommutator
and Poisson bracket,
\begin{equation}\label{eq:pos2}
\int d^3x d^3y A_i^{\alpha\,u}(\vec x) A_j^{*\beta\,v}(\vec y) \{\Psi_i^{\alpha\,u}(\vec x),\Psi_j^{\dagger \, \beta\,v }(\vec y)\}_C= \int d^3x d^3y  A_i^{\alpha\,u}(\vec x) A_j^{*\beta\,v}(\vec y) i[\Psi_i^{\alpha\,u}(\vec x),\Psi_j^{\dagger \, \beta\,v }(\vec y)]_C~~~.
\end{equation}\label{eq:pos3a}
From $\Psi_j^{\dagger \, \beta \,v}= i  P_j^{\beta\,v}-\epsilon_{jkl} P_k^{\delta\,v} \sigma_l^{\delta \beta}$,
we find that
\begin{align}\label{eq:pos3}
[\Psi_i^{\alpha\,u}(\vec x),\Psi_j^{\dagger \, \beta \,v}(\vec y)]_C=&-i\big(\delta_{ij}\delta^{\alpha\beta}+i\epsilon_{jik} \sigma_k^{\alpha \beta}\big)\delta^{uv}\delta^3(\vec x-\vec y)\cr
 =& -i (\sigma_j\sigma_i)^{\alpha\beta}\delta^{uv}\delta^3(\vec x-\vec y)
= -2i (\delta_{ij}-\frac{1}{2}\sigma_i\sigma_j)^{\alpha\beta}\delta^{uv}\delta^3(\vec x-\vec y)~~~.\cr
\end{align}
Multiplying by $i/2$, and writing $A_i^{\alpha\,u}=R_i^{\alpha\,u}+iI_i^{\alpha\,u},\, i=1,2,3,\,\alpha=1,2,\,u=1,...,n$, with $R$ and $I$ real, the right hand side of  Eq. \eqref{eq:pos2} evaluates to \big(we suppress the internal symmetry index $u$ from here on, so $(R_i^{\alpha})^2$ means
$\sum_{u=1}^n(R_i^{\alpha\,u})^2$ , etc.\big)
\begin{equation}\label{eq:pos4}
\sum_{i=1}^3\sum_{\alpha=1}^2  \big( (R_i^{\alpha})^2+(I_i^{\alpha})^2\big) - \frac{1}{2}\big((R_2^1-I_1^1+I_3^2)^2+(R_1^1+I_2^1-R_3^2)^2+(R_2^2+I_1^2+I_3^1)^2+(R_1^2-I_2^2+R_3^1)^2\big)~~~.
\end{equation}
If all three components $A_i^{\alpha},\, i=1,...,3$ are present, the expression in Eq. \eqref{eq:pos4} is {\it not} positive semidefinite.  But when only two of the three components are present, as a result of application of a constraint, then each of the four squared terms on the right hand
side of Eq. \eqref{eq:pos4} contains only two terms, and so the expression in Eq. \eqref{eq:pos4} is positive semidefinite by virtue of the inequality
\begin{equation}\label{eq:posineq}
X^2+Y^2-\frac{1}{2}(X\pm Y)^2= \frac{1}{2}(X\mp Y)^2 \geq 0~~~.
\end{equation}
Another way of seeing this, noted by both Velo and Zwanziger \cite{velo} and Allcock and Hall \cite{allcock}, is that because $\sum_{i=1}^3 \sigma_i \sigma_i =3$, the
expression $W_{ij}=\delta_{ij}- \frac{1}{2} \sigma_i\sigma_j$ is not a projector.  But when one component of $\vec \sigma$, say $\sigma_3$,  is replaced by 0, so that one has  $\sum_{i=1}^3 \sigma_i \sigma_i =\sum_{i=1}^2 \sigma_i \sigma_i =2$,
then
\begin{equation}\label{eq:wproj}
\sum_l W_{il}W_{lj}=\delta_{ij}- 2 \frac{1}{2}\sigma_i\sigma_j+\frac{1}{4} \sigma_i \sum_{l=1}^2 \sigma_l \sigma_l \sigma_j = \delta_{ij}-\frac{1}{2}\sigma_i\sigma_j= W_{ij}~~~,
\end{equation}
and $W_{ij}$ is a projector and hence is positive semidefinite.  So we anticipate that proving positivity will require projection of
Eq. \eqref{eq:pos3} into a subspace obeying  at least one constraint on $\vec \Psi$.

The next step is to use the  property that the Dirac bracket of linear quantities $F$ and $G$ reduces to
the canonical bracket of their projections into the subspace obeying the constraints,
when (as is the case here) all constraints are second class, that is they all appear in the Dirac bracket \cite{hanson}.
Referring to Eq. \eqref{eq:fullbrac}, let us define
\begin{align}\label{eq:fproj}
\tilde F=&F-\sum_a \sum_b [F,\Phi_a]_C T^{-1}_{ab} \Phi_b~~~,\cr
\tilde G=&G-\sum_a \sum_b [G,\Phi_a]_C T^{-1}_{ab} \Phi_b~~~,\cr
 \end{align}
so that
\begin{align}\label{eq:tildebrac}
[\tilde F,\Phi_c]_C=&[F,\Phi_c]_C- \sum_a\sum_b [F,\Phi_a]_C T^{-1}_{ab} [\Phi_b,\Phi_c]_C  ~~~\cr
=&[F,\Phi_c]_C- \sum_a\sum_b [F,\Phi_a]_C T^{-1}_{ab}T_{bc}  ~~~\cr
=&[F,\Phi_c]_C- \sum_a [F,\Phi_a]_C \delta_{ac}=0 ~~~,\cr
\end{align}
and similarly for $\tilde G$.  As a result of this relation, which holds when the canonical brackets
are simply numbers (as in the case here where $\Phi_c$ and $F,\,G$ are linear),
together with symmetry  of the canonical bracket $[\tilde G, \Phi_c]_C=[\Phi_c, \tilde G]_C$, we see that
\begin{equation}\label{eq:tildebrac1}
[ F, G]_D= [\tilde F,\tilde G]_C~~~.
\end{equation}
These properties of Eqs. \eqref{eq:fproj}--\eqref{eq:tildebrac1} carry over when we replace Grassmann numbers with operators, and  classical brackets with anticommutators, since
in the linear case all anticommutators of linear quantities are c-numbers that commute with the operators, and since the
anticommutator is symmetric.  Thus we have
\begin{equation}\label{eq:tildebrac3}
  \{ \Psi_i^{\alpha}(\vec x), \Psi_j^{\dagger \, \beta}(\vec y\}_D=   \{\tilde \Psi_i^{\alpha}(\vec x),\tilde \Psi_j^{\dagger \, \beta}(\vec y)\}_C~~~.
\end{equation}

To further study the properties of $\tilde \Psi_i(\vec x)$ and $\tilde
\Psi_j^{\dagger}(\vec y)$  (with spinor indices suppressed), let us now return to our
original labeling of the constraints by $\phi_a$ and $\chi_a$ as in Eq. \eqref{eq:simplebrac}, so that we have
in the Dirac bracket formalism
\begin{equation}\label{eq:originaltilde}
\tilde  \Psi_i(\vec x)=\Psi_i(\vec x)-\sum_a \sum_b [ \Psi_i(\vec x),\chi_a]_C M_{ab}^{-1} \phi_b~~~,
\end{equation}
and a similar equation (with the roles of $\phi_a$ and $\chi_a$ interchanged) for $\tilde \Psi_j^{\dagger}(\vec y)$, with $a,b$ summed from 3 to 4.  We now note two important
properties of this equation.  The first is that it is invariant under replacement of the constraints $\chi_a$ by
any linear combination $\chi_a^{\prime}= \chi_b K_{ba}$, with the matrix $K$ nonsingular, since the factors
$K$ and $K^{-1}$ cancel between $\chi_a^{\prime}$ and $M_{ab}^{\prime\,-1}$.  (More generally, the Dirac bracket is invariant under
replacement of the constraints by any nonsingular linear combination of the constraints, reflecting the fact that the
Dirac bracket is a projector onto the subspace obeying the constraints, and this subspace is invariant
under replacement of the constraints by any nonsingular linear combination of the constraints.) The second is that
if we act on $\tilde \Psi_i(\vec x)$ with either $D_{\vec x\,i}$ or $(\vec \sigma \times D_{\vec x})_i$, we get zero.
For example, recalling that in covariant radiation gauge $D_{\vec x\,i} \Psi_i(\vec x)=\phi_4(\vec x)$, we have (with spatial variable labels
$\vec x$ suppressed)
\begin{equation}\label{eq:dgives0}
D_i\tilde  \Psi_i=\phi_4 -\sum_a \sum_b [ \phi_4,\chi_a]_C M_{ab}^{-1} \phi_b
=\phi_4 -\sum_a \sum_b M_{4a}M_{ab}^{-1} \phi_b=\phi_4 -\sum_b \delta_{4b} \phi_b=0~~~,
\end{equation}
and similarly for  $(\vec \sigma \times D_{\vec x})_i$,  with $\phi_4$ replaced by $\phi_3$.

Let us now write $\tilde  \Psi_i(\vec x)$ as a projector $R_{ij}(\vec x,\vec y)$ acting on $\Psi_j(\vec y)$,  giving
after an integration by parts on $\vec y$,
\begin{align}\label{eq:projectordef}
\tilde  \Psi_i(\vec x)=&\int d^3y R_{ij}(\vec x,\vec y) \Psi_j(\vec y)~~~,\cr
R_{ij}(\vec x,\vec y)=&\delta_{ij}\delta^3(\vec x-\vec y)+ \sum_a \sum_b \int d^3z [ \Psi_i(\vec x),\chi_a(\vec z)]_C M_{ab}^{-1}(\vec z,\vec y)  \overleftarrow{\eta}_{b\,j}(\vec y)~~~,\cr
\end{align}
with
\begin{equation}\label{eq:etadef}
\overleftarrow{\eta}_{3\,j}(\vec y)=(\vec \sigma \times \overleftarrow{D}_{\vec y})_j~~~,~~
\overleftarrow{\eta}_{4\,j}(\vec y)=\overleftarrow{D}_{\vec y\,j}~~~.
\end{equation}
By virtue of Eq. \eqref{eq:dgives0} and its analog for $\vec \sigma \times \vec D$, we have
\begin{align}\label{eq:dgives01}
D_{\vec x\,i} R_{ij}(\vec x,\vec y)=&0~~~,\cr
(\vec \sigma \times \vec D_{\vec x})_i R_{ij}(\vec x,\vec y)=&0~~~.\cr
\end{align}
Since
\begin{equation}\label{eq:newsigmaid}
\vec \sigma \cdot \vec D_{\vec x} \sigma_{i} R_{ij}(\vec x,\vec y)=D_{\vec x\,i} R_{ij}(\vec x,\vec y)+i (\vec \sigma \times \vec D_{\vec x})_i R_{ij}(\vec x,\vec y)~~~,
\end{equation}
then assuming  that $\vec \sigma \cdot \vec D$ is invertible Eqs. \eqref{eq:dgives01} also imply that
\begin{equation}\label{eq:sigmagives0}
\sigma_i R_{ij}(\vec x,\vec y)=0~~~.
\end{equation}
Next let us focus on the bracket $[ \Psi_i(\vec x),\chi_a(\vec z)]_C $ appearing as the first factor inside the sum.
Setting $\vec L=\vec D$  in Eq. \eqref{eq:neededbracs} we have
\begin{align}\label{eq:neededbracs1}
[\vec \Psi(\vec x),\chi_3(\vec z)]_C=&2 \vec D_{\vec x} \delta^3(\vec x-\vec z)~~~,\cr
[\vec \Psi(\vec x),\chi_4(\vec z)]_C=&(i \vec D_{\vec x}- \vec \sigma \times \vec D_{\vec x})\delta^3(\vec x-\vec z)~~~.\cr
\end{align}
Using the invariance of $\tilde \Psi_i$, or equivalently of $R_{ij}$, under replacement of $\chi_3,\,\chi_4$ by any nondegenerate
linear combination of  $\chi_3,\,\chi_4$,  let us choose the new combinations so that
\begin{align}\label{eq:neededbracs2}
[\vec \Psi (\vec x),\chi_3(\vec z)]_C=& (\vec \sigma \times \vec D_{\vec x})\delta^3(\vec x-\vec z)=\vec \eta_3(\vec x)\delta^3(\vec x-\vec z)~~~,\cr
[\vec \Psi(\vec x),\chi_4(\vec z)]_C=& \vec D_{\vec x}\delta^3(\vec x-\vec y)=\vec \eta_4(\vec x)\delta^3(\vec x-\vec z) ~~~.\cr
\end{align}
Substituting this into Eq. \eqref{eq:projectordef}, we get the symmetric  expression
\begin{equation}\label{eq:projectsym}
R_{ij}(\vec x,\vec y)=\delta_{ij}\delta^3(\vec x-\vec y)+ \sum_a \sum_b \int d^3z \vec \eta_{a\,i}(\vec x)  M_{ab}^{-1}(\vec x,\vec y)  \overleftarrow{\eta}_{b\,j}(\vec y)~~~.
\end{equation}
By virtue of this symmetry, the projector $R_{ij}$ is annihilated by the constraints $\overleftarrow{D}_{\vec y\,j}$ and
$(\vec \sigma \times \overleftarrow{D}_{\vec y})_j$ acting from the right, which in turn implies that
in addition to  Eq. \eqref{eq:sigmagives0} we also have
\begin{equation}\label{eq:sigmagives01}
 R_{ij}(\vec x,\vec y)\sigma_j=0~~~.
\end{equation}
An explicit construction of $R_{ij}(\vec x,\vec y)$ and verification of Eqs. \eqref{eq:sigmagives0}    and  \eqref{eq:sigmagives01}   is given in Appendix C.

Returning now to Eqs. \eqref{eq:pos1} and \eqref{eq:tildebrac3}, writing $\tilde \Psi_i^{\alpha}$ and $\tilde \Psi^{\dagger\,\beta}_j$ in terms of  projectors acting
on $\Psi_i^{\alpha}$ and $\Psi^{\dagger\,\beta}_j$, we have  (using $\sigma_m^{\epsilon \delta}=\sigma_m^{*\delta \epsilon}$, and continuing
to suppress internal symmetry indices $u,v$, which are contracted in the same pattern as the spatial vector and spin indices)
\begin{align}\label{eq:tildebrac4}
&\int d^3x  \int d^3y A_i^{\alpha}(\vec x) A_j^{*\, \beta}(\vec y) \{\Psi_i^{\alpha}(\vec x), \Psi_j^{\dagger\, \beta}(\vec y)\}_D \cr= &\int d^3x \int d^3y A_i^{\alpha}(\vec x)A_j^{*\,\beta}(\vec y)  \{\tilde \Psi_i^{\alpha}(\vec x),\tilde \Psi_j^{\dagger\,\beta}(\vec y)\}_C\cr= &
\int d^3x \int d^3y A_i^{\alpha}(\vec x)A_j^{*\beta}(\vec y) \int d^3z \int d^3w \,R_{il}^{\alpha\gamma}(\vec x,\vec z)\{\Psi_l^{\gamma}(\vec z), \Psi_m^{\dagger\,\delta}(\vec w)\}_C R^{*\beta\delta}_{jm}(\vec y,\vec w)\cr
=&\int d^3x \int d^3y A_i^{\alpha}(\vec x)A_j^{*\beta}(\vec y) \int d^3z \int d^3w\, R_{il}^{\alpha\gamma}(\vec x,\vec z)2\left(\delta_{lm}\delta^{\gamma\delta}-\frac{1}{2}\sigma_{l}^{\gamma\epsilon}\sigma_m^{*\delta\epsilon}\right)\delta^3(\vec z-\vec w) R^{*\beta\delta}_{jm}(\vec y,\vec w)\cr
=&2\int d^3z \Big[\int d^3x  A_i^{\alpha} (\vec x)R_{il}^{\alpha\gamma}(\vec x,\vec z)\Big]
\Big[ \int d^3y A_j^{\beta}(\vec y) R^{\beta\gamma}_{jl}(\vec y,\vec z)\Big]^*~~~,\cr
\end{align}
which is  positive semidefinite.

We conclude that the anticommutator of $\vec \Psi$ with $\vec \Psi^{\dagger}$ is manifestly positive semidefinite in covariant radiation gauge.  The duality of the $\phi_{3,4}$ and $\chi_{3,4}$ constraints in this gauge is essential to reaching this conclusion; if gauge fixing were omitted,
or if another gauge were chosen, this symmetry would not be present and we could not deduce positivity in a similar fashion.

\section{Alternative Lagrangian and Hamiltonian for the $\vec \Psi$ equation in covariant radiation gauge}

Up to this point we have worked with the original action of Eq. (I-23) and the canonical momentum derived from it.  We give
here another approach, based on setting up an action for the $\vec \Psi$ equation of motion from which $\Psi_0$ has been
eliminated by the secondary constraint,
\begin{align}\label{eq:neweqmo}
D_0 \vec \Psi=& \vec D \vec R \cdot \vec \Psi +  i \vec D \times \vec \Psi~~~,\cr
\vec R =& (\vec \sigma \cdot \vec B) ^{-1} (\vec B + \vec \sigma \times \vec E)~~~,
\end{align}
which holds when the primary constraint $\chi=0$ is obeyed.
Consider the self-adjoint action
\begin{equation}\label{eq:newaction}
\hat S=\int d^3x \hat L =\frac{i}{2}\int d^4x \vec{\Psi}^{\dagger} \cdot \big( D_0 \vec \Psi -i \vec D \times \vec \Psi
- \vec D \vec R \cdot \vec \Psi - \vec{R}^{\dagger} \vec D \cdot \vec \Psi\big)~~~.
\end{equation}
Varying with respect to $\vec \Psi^{\dagger}$, and imposing two constraints:  (i) the primary constraint
$\chi=\vec \sigma \cdot \vec D \times \vec \Psi=0$, and (ii) the gauge fixing constraint $\vec D \cdot \vec \Psi=0$,
we get the equation of motion of Eq. \eqref{eq:neweqmo}.  For the canonical momentum conjugate to $\vec \Psi$, we
find
\begin{equation}\label{eq:newcanmon}
\vec P=\frac{\partial^L \hat S}{\partial(\partial_0\vec\Psi)}= -\frac{i}{2} \vec{\Psi}^{\dagger}~~~,
\end{equation}
which implies that
\begin{equation}\label{eq:canmominv}
 \vec{\Psi}^{\dagger}=2i\vec P~~~.
\end{equation}
For the Hamiltonian corresponding to the new action, we find (again for simplicity taking $A_0=0$, and integrating the middle
term by parts)
\begin{align}\label{eq:newham}
\hat H= &\int d^3x \partial_0 \vec \Psi \cdot \vec P - \hat L  \cr
=& \frac{1}{2} \int d^3x \vec {\Psi}^{\dagger} \cdot \big( -\vec D \times \vec \Psi - i \overleftarrow{D} \vec R \cdot \vec \Psi
+ i \vec{R}^{\dagger} \vec D \cdot \vec \Psi \big) \cr
=&-\frac{1}{2} \int d^3x \vec{\Psi}^{\dagger} \cdot \vec D \times \vec \Psi~~~, \cr
\end{align}
where in going from the second to the final line we have used the constraint $\vec D \cdot \vec \Psi=0$ and its adjoint $\vec{\Psi}^{\dagger} \cdot
\overleftarrow{D}=0$.  The Hamiltonian $\hat H$ is again the energy integral calculated from the left chiral part of the stress-energy tensor,
and expressed in terms of the canonical momentum is
\begin{equation}\label{eq:newham1}
\hat H= -i \int d^3x \vec P \cdot \vec D \times \vec \Psi~~~.
\end{equation}

From here on the argument parallels that of Secs. 2 and 3, but is simpler.  For the canonical bracket of $\Psi_i(\vec x)$ with
$\Psi_j(\vec y)$ we have
\begin{equation}\label{eq:newcanbrac}
[\Psi_i(\vec x),\Psi^{\dagger}_j(\vec y)]_C= [\Psi_i(\vec x),2iP_j(\vec y)]_C
= -2i \delta_{ij}\delta^3(\vec x-\vec y)  ~~~,
\end{equation}
and so multiplying by $i$ to convert to a canonical anticommutator we get
\begin{equation}\label{eq:newcananti}
\{\Psi_i(\vec x),\Psi^{\dagger}_j(\vec y)\}_C= 2 \delta_{ij}\delta^3(\vec x-\vec y) ~~~,
\end{equation} which is positive semidefinite. The complete set of constraints is
\begin{align}\label{eq:newconstraints}
\phi_3= &\chi=\vec \sigma \cdot \vec D \times \vec \Psi~~~,\cr
\phi_4=&\vec D \cdot \vec \Psi~~~,\cr
\chi_3=& \chi^{\dagger}=-\vec{\Psi}^{\dagger} \times \overleftarrow{D} \cdot \vec \sigma~~~
=2i\vec P\cdot \vec \sigma \times  \overleftarrow{D},\cr
\chi_4=&  \vec{\Psi}^{\dagger} \cdot  \overleftarrow{D}=2i \vec P \cdot \overleftarrow{D}~~~.\cr
\end{align}
The constraints $\phi_3, \, \phi_4$ are identical to $\phi_1,\,\phi_2$ of Eq. \eqref{c1}, while the constraints $\chi_3, \, \chi_4$
are  $\chi_1, \, \chi_2$ of Eq. \eqref{c1} up to an invertible linear transformation (just interchange of the $\chi$ constraints and division by $2i$).   Thus the projector $R_{ij}(\vec x,\vec y)$ is the
same as that calculated in Appendix A, and the Dirac anticommutator given by Eq. \eqref{eq:tildebrac3} is is positive
semidefinite by Eq. \eqref{eq:tildebrac4}, this time without using the fact that $R_{ij}$ is projected to zero by $\sigma_i$ and $\sigma_j$.

\section{Lorentz covariance of covariant radiation  gauge and Lorentz invariance of the Dirac bracket}

We study next the behavior of covariant radiation gauge and the Dirac bracket under Lorentz boosts.  The Rarita-Schwinger field
$\psi_{\mu}^{\alpha}$ and its left-handed chiral projection $\Psi_{\mu}^{\alpha}$ both have a four-vector index $\mu$ and
a spinor index $\alpha$.  Under an infinitesimal Lorentz transformation, the transformations acting on these two types of indices
are additive, and so can be considered separately.  The spinor indices are transformed as in the usual spin $\frac{1}{2}$ Dirac
equation by a matrix constructed from the Dirac gamma matrices, which commutes with $D_{\mu}$.  Hence the spinor index  transformation leaves
the covariant radiation gauge condition $\vec D \cdot \vec \Psi$ invariant.

This leaves the transformation on the vector index to be considered, and this is a direct analog of the Lorentz transformation of radiation
gauge in quantum electrodynamics \cite{zumino}.  Since the radiation gauge condition is invariant under spatial rotations, we
only have to consider a Lorentz boost,
\begin{align}\label{eq:boost1}
\vec x \to &\vec x^{\,\prime}= \vec x + \vec v t~~~, \cr
x^0=&t \to  t^{\prime} = t+ \vec v \cdot \vec x~~~.\cr
\end{align}
Under this boost, the field $\vec \Psi$ transforms as
\begin{equation} \label{eq:boost2}
\vec \Psi \to \vec \Psi^{\prime} = \vec \Psi + \vec v \Psi^0~~~.
\end{equation}
For an observer in the boosted frame, covariant radiation gauge would be
$\vec D_{\vec x^{\prime}} \cdot \vec \Psi^{\prime}=0$, with
$\vec D_{\vec x^{\prime}}= \vec \nabla_{\vec x^{\prime}}+ g \vec A^{\prime}$, where $\vec A^{\prime}=\vec A + O(\vec v)$.
Applying this to $\vec \Psi^{\prime}(\vec x^{\prime},t^{\prime})$
and using the covariant radiation gauge condition in the initial frame, we get
\begin{equation} \label{eq:vecd1}
\vec D_{\vec x^{\prime}}\cdot  \vec \Psi^{\prime}
= v_j \Sigma_j(\vec x,t)~~~,
\end{equation}
with $\Sigma_j(\vec x,t)$ a local polynomial in $\vec \Psi,~\Psi_0$ and the gauge fields,
where we have dropped primes on the right hand side since there is an explicit factor of $\vec v$.
So in the boosted frame $\vec \Psi^{\prime}$ does not obey the covariant radiation gauge condition, but this
can be restored by making a gauge transformation
\begin{equation}\label{eq:gaugerestore}
\vec \Psi^{\prime} \to \vec \Psi^{\prime} -\vec D
(\vec D^2)^{-1} v_j \Sigma_j(\vec x,t)~~~.
\end{equation}
Hence the covariant radiation gauge condition is Lorentz boost covariant, although not Lorentz boost invariant.

Referring now to Eq. \eqref{c10}, we note that the covariant radiation gauge Dirac bracket and the anticommutation relations are invariant
under infinitesimal Rarita-Schwinger gauge transformations, such as that of Eq. \eqref{eq:gaugerestore}, up to a
remainder that is quadratic in the gauge parameter.   Hence
the covariant radiation gauge Dirac bracket and the anticommutation relations following from it are Lorentz invariant, since a finite Lorentz
transformation can be built up from a series of infinitesimal ones.

\section{Path integral quantization}

An alternative method of quantization to the Dirac bracket approach is setting up a Feynman path integral.  Again, we will specialize to the case where the external gauge potentials, and hence $\vec D$, are time independent, since the simplest discussions
of path integrals for  constrained systems assume time-independent constraints.  As noted above, this assumption can be dropped when the gauge field is quantized along with the Rarita-Schwinger field, leading to a  more complex system of constraints and constraint brackets.

When the constraints are time independent, the classical brackets of Eqs. (I-57) and (I-58)
 have the form needed to apply the Faddeev-Popov  \cite{faddeev} method for path integral
quantization. (This has been applied in the free Rarita-Schwinger case by Das and Freedman \cite{dasb} and by Senjanovi\'c \cite{sejn}.) The general formula of \cite{faddeev} for the in to out $S$ matrix element (up to a constant proportionality factor) reads
\begin{align}\label{eq:smatrix}
\langle {\rm out}|S| {\rm in} \rangle  \propto &\int \exp\big(iS(q,p)\big) \prod_t d\mu\big(q(t),p(t)\big)~~~,\cr
d\mu(q,p)=&\prod_a \delta(\chi_a)\delta(\phi_a) (\det [\phi_a,\chi_b])^\xi \prod_i dp_i dq_i~~~,\cr
\end{align}
where $\xi=1$ when all canonical variables are bosonic, and $\xi=-1$ in our case in which all canonical variables
are fermionic, or Grassmann odd.  In applying this formula, we note that since the action $S$ of Eq. (I-23)
and the bracket matrix $M$ of Eqs. (I-59)-(I-62)  are
 independent of $P_{\Psi_0}$ and $P_{\Psi_0^{\dagger}}$, we can immediately integrate out the delta functions in these
two constraints.  Also, since the canonical momentum $\vec P$ is related to $\vec{\Psi}^{\dagger}$ by the constant numerical
transformation of Eq. (I-54), we can take $\vec{\Psi}^{\dagger}$ as the integration variable instead of
$\vec P$, up to an overall proportionality constant.  So we have  the formula, after an integration by parts in the second term,
\begin{align}\label{eq:smatrix1}
\langle {\rm out}|S| {\rm in} \rangle\propto & \int \exp\big(i\frac{1}{2}\int d^4x  [-\Psi_{0}^{\dagger} \vec \sigma \cdot \vec D \times \vec {\Psi}-\vec {\Psi}^{\dagger} \cdot \vec \sigma \times \overleftarrow {D} \Psi_{0}
+\vec{\Psi} ^{\dagger} \cdot \vec D \times \vec \Psi - \vec{\Psi}^{\dagger} \cdot \vec \sigma \times D_{0} \vec{\Psi}]\big) \cr
\times&
\prod_{t,\vec x} d\mu\big(\Psi_0,\Psi_0^{\dagger},\vec \Psi, \vec \Psi^{\dagger}\big)\cr
=&
 \int \exp\big(i\frac{1}{2}\int d^4x  [-\Psi_{0}^{\dagger}\chi -\chi^{\dagger} \Psi_{0}
+\vec{\Psi} ^{\dagger} \cdot \vec D \times \vec \Psi - \vec{\Psi}^{\dagger} \cdot \vec \sigma \times D_{0} \vec{\Psi}]\big) \cr
\times&
\prod_{t,\vec x} d\mu\big(\Psi_0,\Psi_0^{\dagger},\vec \Psi, \vec \Psi^{\dagger}\big)
~~~.\cr
\end{align}
Here
\begin{equation}\label{eq:measure}
d\mu\big(\Psi_0,\Psi_0^{\dagger},\vec \Psi, \vec \Psi^{\dagger}\big)
=\left(\prod_{c=2}^4 \delta(\chi_c)\delta(\phi_c)\right) (\det [\phi_a,\chi_b])^{-1} d\Psi_0 d\Psi_0^{\dagger} d\vec \Psi d\vec \Psi^{\dagger}~~~,
\end{equation}
with $d\Psi_0$ and $d\Psi_0^{\dagger}$ each a product over the spinor components, and
$d\vec \Psi$ and $d\vec \Psi^{\dagger}$ each a product over the spinor-vector components.

As our next step, we can carry out the integrations over $\Psi_0$ and $\Psi_0^{\dagger}$, using the delta functions $\delta(\phi_2)$ and $\delta(\chi_2)$ .  This leaves the
formula
\begin{align}\label{eq:smatrix2}
\langle {\rm out}|S| {\rm in} \rangle\propto& \int \exp\big(i\frac{1}{2}\int d^4x  [-\vec{\Psi}^{\dagger}
\cdot (\vec B + \vec \sigma \times \vec E) (\vec \sigma \cdot \vec B)^{-1} \chi \cr
-&\chi^{\dagger}(\vec \sigma \cdot \vec B)^{-1}(\vec B + \vec \sigma \times \vec E) \cdot \vec \Psi
+\vec{\Psi} ^{\dagger} \cdot \vec D \times \vec \Psi - \vec{\Psi}^{\dagger} \cdot \vec \sigma \times D_{0} \vec{\Psi}]\big) \cr
\times&
\prod_{t,\vec x} d\mu\big(\vec \Psi, \vec \Psi^{\dagger}\big)~~~,\cr
\end{align}
with
\begin{equation}\label{eq:measure2}
d\mu\big(\vec \Psi, \vec \Psi^{\dagger}\big)
=\left(\prod_{c=3}^4 \delta(\chi_c)\delta(\phi_c)\right) (\det [\phi_a,\chi_b])^{-1} d\vec \Psi d\vec \Psi^{\dagger}~~~,
\end{equation}
so that only the remaining constraints $\phi_{3,4},\,\chi_{3,4}$ are  used in constructing the determinant $\det [\phi_a,\chi_b]$.

Finally, using the delta functions $\delta(\phi_3)=\delta(\chi)$ and $\delta(\chi_3)=\delta(\chi^{\dagger})$ to simplify the exponent,
 we end up with the elegant formula
\begin{align}\label{eq:smatrix4}
\langle {\rm out}|S| {\rm in} \rangle\propto & \int \exp\big(i\frac{1}{2}\int d^4x
\vec {\Psi}^{\dagger} \cdot[ \vec D \times \vec \Psi - \vec \sigma \times D_{0} \vec{\Psi}]\big) \cr
\times&\prod_{t,\vec x} d\mu\big(\vec \Psi, \vec \Psi^{\dagger}\big)~~~,\cr
\end{align}
which as in Dirac bracket quantization, employs as Hamiltonian the energy integral computed in Eq. (I-35) from the stress-energy tensor.
In using  this formula, the customary procedure \cite{fradkin} would be to put the bracket matrix that is the argument of the determinant back into the exponent by introducing bosonic ghost fields $\phi_G$.

\section{Conclusion and discussion}

To conclude, we see that when a covariant radiation gauge constraint is included, the problems with canonical quantization found
in \cite{johnson} and \cite{velo} are avoided:  The Dirac bracket is well-defined in the limit of zero external fields, and
is positive semidefinite.  Thus our conclusion in (I) that the classical theory  of gauged Rarita-Schwinger fields is consistent extends to
 the quantized theory of gauged Rarita-Schwinger fields as well.   As noted in (I), this means that in constructing grand unified theories, one can contemplate an anomaly cancellation mechanism in which the gauge anomalies of Rarita-Schwinger fields cancel against those
of spin-$\frac{1}{2}$ fields, as first suggested in \cite{marcus} and as used in the $SU(8)$ family unification model of \cite{adler}.

Some final remarks:

\begin{enumerate}

\item  In quantizing, we assumed that the gauge fields $\vec A$  are time independent, so that $d/dt$ and $\vec D$ commute.  As noted,
this assumption can be dropped if the gauge fields are treated as dynamical variables, leading to an extension of the bracket structure, involving fermionic brackets as well as bosonic ones.  (For a discussion of bosonic versus fermionic constraints, see \cite{junker}.)

\item  In demonstrating positivity of the anticommutator in Sec. 3 (but not in Sec. 4), we used the condition $\vec \sigma \cdot \vec \Psi=0$. Deriving this
from the covariant radiation gauge condition $\vec D \cdot \vec \Psi=0$ assumed the invertibility of $\vec \sigma \cdot \vec D$, and attainability of covariant radiation gauge assumed the
invertibility of $(\vec D)^2$.  The conditions for invertibility of these two operators remain to be studied.
(The open space index theorems of Callias \cite{callias}  and
Weinberg \cite{callias} involve $\vec \sigma \cdot \vec D + i\phi$, with $\phi$ a scalar field, and so do not give information about the invertibility of $\vec \sigma \cdot \vec D$.)

\end{enumerate}

\section{Acknowledgements}

I wish to thank Edward Witten for  conversations about gauging Rarita-Schwinger fields and Rarita-Schwinger scattering from photons,
among other topics.   I also wish to acknowledge the  various people who
asked about the status of gauged Rarita-Schwinger fields  when I gave seminars on \cite{adler}.  Following on the initial draft
of this paper, I had a fruitful correspondence with Stanley Deser and Andrew Waldron about gauge invariance and counting degrees
of freedom when invariance of the action is conditional on a constraint.   I  wish to thank Thomas Spencer for a very helpful conversation which emphasized the significance of the gauge invariants, and Laurentiu Rodina for an explication of the paper \cite{mcgady} that uses
  ``on-shell'' methods.  This work was supported in part by the National Science Foundation under Grant
No. PHYS-1066293 through the hospitality of the Aspen Center for Physics.

\appendix
\section{Construction of the projector $R_{ij}(\vec x,\vec y)$}

Since there are only two $\phi_{a}$ constraints and two $\chi_{a}$ constraints, we index them $a=1,2$ rather than $a=3,4$ as in the text, and use
the invariance of $R_{ij}(\vec x,\vec y)$ under changing the linear combination of the $\chi_{a}$ constraints.
We start from the constraint set
\begin{align}\label{c1}
\phi_1=\vec \sigma \times \vec D \cdot \vec \Psi~~,~~~\chi_1= \vec P \cdot \overleftarrow{D}~~~,\cr
\phi_2=\vec D \cdot \vec \Psi~~,~~~\chi_2= \vec P \cdot \vec \sigma \times \overleftarrow{D}~~~.\cr
\end{align}
For the bracket matrix
\begin{equation}\label{c2}
M_{ab}(\vec x, \vec y)= [\phi_a(\vec x),\chi_b(\vec y)]_C= \left( \begin{array} {cc}
 \hat{\cal A}&\hat{\cal B} \\
 \hat{\cal C}&\hat{\cal D}\\ \end{array} \right)~~~,
\end{equation}
we find the matrix elements
\begin{align}\label{c3}
\hat {\cal A}=& -ig  \vec \sigma \cdot \vec B \delta^3(\vec x-\vec y)    ~~~,\cr
\hat {\cal B}=&  \big(2 (\vec D_{\vec x})^2 + g  \vec \sigma \cdot \vec B \big)  \delta^3(\vec x-\vec y)
=\delta^3(\vec x-\vec y) \big(2  (\overleftarrow{D}_{\vec y})^2  + g  \vec \sigma \cdot \vec B \big)     ~~~,\cr
\hat {\cal C}=&  (\vec D_{\vec x})^2     \delta^3(\vec x-\vec y) =  \delta^3(\vec x-\vec y) (\overleftarrow{D}_{\vec y})^2   ~~~,\cr
\hat {\cal D}=&    ig  \vec \sigma \cdot \vec B \delta^3(\vec x-\vec y)    ~~~.\cr
\end{align}
We write the inverse matrix $M^{-1}(\vec z,\vec w)$ as
\begin{equation}\label{c4}
\left( \begin{array} {cc}
 \hat{\cal F}&\hat{\cal G} \\
 \hat{\cal H}&\hat{\cal I}\\ \end{array} \right)~~~,
 \end{equation}
which obeys
\begin{equation}\label{c5}
\left( \begin{array} {cc}
 \hat{\cal A}&\hat{\cal B} \\
 \hat{\cal C}&\hat{\cal D}\\ \end{array} \right)\left( \begin{array} {cc}
 \hat{\cal F}&\hat{\cal G} \\
 \hat{\cal H}&\hat{\cal I}\\ \end{array} \right)
 =\left( \begin{array} {cc}
 \hat{\cal F}&\hat{\cal G} \\
 \hat{\cal H}&\hat{\cal I}\\ \end{array} \right)\left( \begin{array} {cc}
 \hat{\cal A}&\hat{\cal B} \\
 \hat{\cal C}&\hat{\cal D}\\ \end{array} \right)=
 \left( \begin{array} {cc}
 1&0 \\
 0&1\\ \end{array} \right)~~~.
 \end{equation}
In terms of the inverse matrix,  the projector $R_{ij}(\vec x,\vec w)$ is given by (with internal symmetry indices suppressed)
\begin{align}\label{c6}
R_{ij}(\vec x,\vec w)=& \delta_{ij} \delta^3(\vec x-\vec w) 1\cr
 +& D_{\vec x \,i}\hat{\cal F}(\vec x-\vec w) (\vec \sigma \times \overleftarrow{D}_{\vec w })_j+ D_{\vec x \,i}\hat{\cal G}(\vec x-\vec w) \overleftarrow{D}_{\vec w \,j}\cr
+&(\vec \sigma \times D_{\vec x })_i \hat{\cal H}(\vec x-\vec w)(\vec \sigma \times \overleftarrow{D}_{\vec w })_j
+(\vec \sigma \times D_{\vec x })_i \hat{\cal I}(\vec x-\vec w)\overleftarrow{D}_{\vec w \,j}~~~.\cr
\end{align}
From this expression, we find
\begin{equation}\label{c7}
D_{\vec x\,i}R_{ij}(\vec x,\vec w)=R_{ij}(\vec x,\vec w)\overleftarrow{D}_{\vec w \,j}
=(\vec \sigma \times D_{\vec x })_i R_{ij}(\vec x,\vec w)=R_{ij}(\vec x,\vec w)(\vec \sigma \times \overleftarrow{D}_{\vec w })_j=0~~~.
\end{equation}
In verifying these, it is not necessary to evaluate the inverse matrix; instead, after contracting on the vector index $i$ or $j$ one
expresses the resulting pre- or post- factor in terms of $\hat{\cal A},...,\hat{\cal D}$ and then uses the algebraic relations following
from multiplying out the matrices in Eq. \eqref{c5}.  Finally, contracting
\begin{align}\label{c8}
\vec \sigma \cdot \vec D_{\vec x} \sigma_i=&(D_{\vec x} + i\vec \sigma \times \vec D_{\vec x})_i~~~,\cr
\sigma_j \vec \sigma \cdot \overleftarrow{D}_{\vec w}=&(\overleftarrow{D}_{\vec w}-i \vec \sigma \times \overleftarrow{D}_{\vec w})_j~~~,\cr
\end{align}
with $R_{ij}(\vec x,\vec w)$, we conclude that
\begin{equation}\label{c9}
\sigma_i R_{ij}(\vec x,\vec y) =R_{ij}(\vec x,\vec y) \sigma_j=0~~~,
\end{equation}
when $\vec \sigma \cdot \vec D$ is invertible.

As a consequence of Eqs. \eqref{eq:projectordef} and  \eqref{c7}, $\tilde \Psi_i(\vec x)$ is invariant under the transformations
\begin{align}\label{c10}
\vec \Psi   \to& \vec \Psi + \vec D \epsilon~~~,\cr
\vec \Psi \to & \vec \Psi + \vec \sigma \times \vec D \epsilon~~~.\cr
\end{align}
The first of these implies that the canonical anticommutation relations are invariant under infinitesimal
Rarita-Schwinger gauge transformations starting from covariant radiation gauge.


\begin{thebibliography}{99}
\bibitem{adlerclassical}  S. Adler, ``Quantized Gauged Massless Rarita-Schwinger Fields'', preceding paper.
\bibitem{adlerarxiv}  S. Adler, ``Classical and Quantum Gauged Massless Rarita-Schwinger Fields'', arXiv:1502.02652 (unpublished).
\bibitem{johnson} K. Johnson and E. C. G. Sudarshan, {\it Ann. Phys.} {\bf 13}, 126 (1961).
\bibitem{velo} G. Velo and D. Zwanziger, Phys. Rev. {\bf 186}, 1337 (1969)
\bibitem{allcock} G. R. Allcock and S. F. Hall, {\it J. Phys. A: Math. Gen.} {\bf 10}, 267 (1977).
\bibitem{hanson} A. Hanson, T. Regge, and C. Teitelboim, ``Constrained Hamiltonian Systems'', Acad. Naz. dei Lincei Volume 373 (1976), pp. 10-11.
\bibitem{zumino}  B. Zumino, {J. Math. Phys.} {\bf 1}, 1 (1960), Eq. (B5).
\bibitem{faddeev} L. D. Faddeev and V. N. Popov, {\it Phys. Lett.} B {\bf 25}, 29 (1967); L. D. Faddeev, Theor. Mat. Fiz.
{\bf 1}, 3 (1969); L. D. Faddeev and A. A. Slavnov, {\it Gauge Fields: Introduction to Quantum Theory}, Benjamin/Cummings (1980), pp. 73-76.
\bibitem{dasb} A. Das and D. Z. Freedman, {\it Nucl. Phys.} B {\bf 114}, 271 (1976).
\bibitem{sejn} G. Senjanovi\'c, {\it Phys. Rev.} D {\bf 16}, 307 (1977).
\bibitem{fradkin} I. A. Batalin and G. A. Vilkovisky, {\it Phys. Lett.} B {\bf 69}, 309 (1977);  E. S. Fradkin and T. E. Fradkina, {\it Phys. Lett.} B {\bf 72}, 343 (1978).
\bibitem{marcus}  N. Marcus, {\it Phys. Lett.} B {\bf 157}, 383 (1985).
\bibitem{adler} S. L. Adler, {\it Int. J. Mod. Phys.} A {\bf 29}, 1450130 (2014).
\bibitem{junker} G. Junker and J. R. Klauder, {\it Eur. J. Phys.} C {\bf 4}, 173 (1998).
\bibitem{callias} C. Callias, {\it Commun. math. Phys.} {\bf 62}, 213 (1978); E. J. Weinberg, {\it Phys. Rev.} D {\bf 20}, 936 (1979).
\bibitem{mcgady}  D. A. McGady and L. Rodina, {\it Phys. Rev.} D {\bf 90}, 084048 (2014).



\end{thebibliography}
\end{document}